# Influence of C/O Ratio on Hot Jupiter Atmospheric Chemistry

Benjamin Fleury*, Murthy S. Gudipati, Bryana L. Henderson, Mark Swain

Science Division, Jet Propulsion Laboratory, California Institute of Technology, 4800 Oak Grove Drive, Pasadena, California 91109, USA

*Corresponding author: benjamin.fleury@jpl.nasa.gov

Pages: 23

Tables: 3

Figures: 6

## Abstract

We have conducted laboratory experiments to study the chemistry in hot Jupiter atmospheres with C/O ratio of 0.35. We have compared our results with the ones obtained previously for atmospheres with a C/O ratio of 1 to investigate the influence of the C/O ratio on the chemistry and formation of photochemical organic aerosol. We found that the C/O ratio and the gas mixture compositions strongly influence the pathways responsible for the formation of $CO_2$. Thermochemical reactions are primarily responsible for the formation of $CO_2$ in low C/O ratio atmospheres, while photochemistry is the dominant process in high C/O ratio atmospheres even if the final $CO_2$ concentration is the same in both cases. Our results show that low C/O atmospheres at the thermochemical equilibrium contain a higher water abundance, while high C/O atmospheres are significantly depleted in water. However, in low C/O atmospheres, the water abundance is not affected by UV photolysis, while our previous work demonstrated that significant amount of water can be produced in high C/O ratio atmospheres. This contrast in water production suggests that photochemistry should be considered when interpreting exoplanet transit spectra. Finally, we did not observe the formation of a detectable amount of non-volatile photochemical aerosols in low C/O atmospheres, in contrast to our previous study. We infer that for C/O ratio < 1, water likely inhibits organic growth and aerosol formation, suggesting that photochemical organic aerosols are likely to be observed in planets presenting a carbon enrichment compared to their host stars.

## 1. Introduction

The atmospheric composition of hot Jupiter exoplanets, Jupiter-sized planets which orbit close to their host stars, have been increasingly studied in the past years with transmission spectroscopy and theoretical modelling. Giant planets are expected to have atmospheres dominated by molecular hydrogen ($H_2$) and helium (He), presumably directly confirmed by the extensive H (Ehrenreich et al. 2015; Vidal-Madjar et al. 2003) and He (Allart et al. 2018; Nortmann et al. 2018; Spake et al. 2018) escape observed for some of these planets. Other chemical species present at lower abundances include carbon monoxide (CO), carbon dioxide ($CO_2$), methane ($CH_4$), and water ($H_2O$) (de Kok et al. 2014; Swain et al. 2009a, 2009b; Tinetti et al. 2007).

Recent observations of transit spectra of hot Jupiter atmospheres show limited spectral modulation due to $H_2O$ that has been largely interpreted as the indicator of the presence of aerosols (Barstow et al. 2016; Iyer et al. 2016; Pinhas et al. 2019; Sing et al. 2016). Whether these aerosols are condensate clouds of photochemical organic aerosols or other refractory materials remains unknown. Although thermochemical equilibrium models predict the formation of condensate clouds with various composition in these hot atmospheres (Lecavelier Des Etangs et al. 2008; Lee et al. 2015; Parmentier et al. 2016), recent laboratory works highlighted that photochemistry could strongly affect the composition of exoplanet atmospheres and lead to the formation of aerosols in a large variety of conditions including the ones encountered in hot Jupiters (Fleury et al. 2019; He et al. 2019, 2018a, 2018b; Hörst et al. 2018). These photochemical aerosols could represent another source of opacity to explain some of the observed transit spectra of hot Jupiter atmospheres, e.g., of HD 189733 b (Lavvas & Koskinen 2017). On the other hand, bulk elemental ratio can also drastically affect the molecular composition of these atmospheres. In the external layers (region with pressure < 1 bar) of atmospheres with temperatures higher than 1000 K, carbon preferentially bonds with oxygen to form CO, and the excess of oxygen bonds with hydrogen to form $H_2O$ when C/O ratio is < 1. At a higher CO ratio ≥ 1, CO remains an abundant species but the water mixing ratio decreases (Drummond et al. 2019; Goyal et al. 2018; Heng & Lyons 2016; Lodders & Fegley 2002; Moses et al. 2013; Tsai et al. 2017; Venot et al. 2015). For this reason, another explanation for the low spectral modulation due to water observed in some hot Jupiter atmospheres is that these atmospheres have low $H_2O$ abundances presumably reflecting high C/O ratios (Madhusudhan 2012; Madhusudhan et al. 2011). However, the existence of such "carbon-rich" exoplanets remains debated. First analysis of the hot Jupiter WASP-12b observations suggested a C/O ratio > 1 (Madhusudhan et al. 2011), but another study found a C/O ratio < 1 using another approach (Kreidberg et al. 2015), leaving the question of the C/O ratio in WASP-12b's atmosphere open. In addition, a recent survey suggests that the carbon enrichment of Hot Jupiter atmospheres compared to their host stars may be common, but uncertainties on C/O measurements in exoplanet atmospheres are large and prevent a firm conclusion from being reached (Brewer et al. 2017).

The C/O ratio varies across exoplanets' host star populations (Brewer et al. 2017; Brewer & Fischer 2016; Delgado Mena et al. 2010) and this variation is likely to be reflected in the composition of exoplanet atmospheres assuming that they are formed with the same materials as the stars. Moreover, various processes in the protoplanetary disks and the planet formation process can affect the exoplanet compositions and have a significant impact on the final C/O ratio (Espinoza et al. 2017; Madhusudhan et al. 2017; Mordasini et al. 2016; Öberg et al. 2011). For these reasons, it is necessary to consider the effects of the C/O ratio on the atmospheric chemistry and the formation of aerosols. Numerous studies have been performed using chemical models (Drummond et al. 2019; Goyal et al. 2018; Heng & Lyons 2016; Madhusudhan 2012; Moses et al. 2013; Tsai et al. 2017; Venot et al. 2015), but corresponding laboratory experiments are still largely nonexistent. Laboratory investigations can provide essential insight into the effects of the C/O ratio on the atmospheric photochemistry and the formation of aerosols. In a previous work, we performed the first laboratory experiments dedicated to the study of the chemistry in hot Jupiter atmospheres (Fleury et al. 2019). This work was focused on the chemistry in atmospheres with T > 1000 K and a C/O ratio of 1 (representing C enhancement compared to the solar value of 0.54), because chemical models predict that the abundances of hydrocarbon and nitrile species increase by several orders of magnitude in these atmospheres compared to atmospheres with a low C/O ratio (Venot et al. 2015). Therefore, they can be considered as better candidates for the formation of complex organic molecules with longer carbon chains. This first study revealed that photochemical aerosols could be produced at temperatures as high as 1500 K and that water could be efficiently formed through photochemical channels. In the present work, we performed new experiments to study the chemistry in hot Jupiter atmospheres at similar temperatures (1173 K – 1473 K) but with lower C/O ratios. We used a gas mixture of $H_2$, $H_2O$, and CO that represents the simplest plausible atmosphere for a hot Jupiter with a C/O ratio < 1. This new study, compared with our previous work, will allow us to assess the evolution of the chemistry in hot Jupiter atmospheres as function of the C/O ratio and atmospheric composition.

## 2. Experimental Setup and Analytical Protocols

## 2.1. Cell for Atmospheric and Aerosol Photochemistry Simulations of Exoplanets (CAAPSE)

We used the CAAPSE experimental setup, which has been described in detail in Fleury et al., (2019) for the studies presented here. Briefly, the cell consists of an alumina tube that is closed at each extremity with an $MgF_2$ window mounted on a stainless-steel flange. The cell is installed inside a customized STT-1600C (SentroTech) oven, which can heat the alumina gas cell up to 1773 K and temperature can be controlled precisely to within a degree or two at the highest temperatures. Before each experiment, the cell was pumped and degassed by heating to and holding at 1473 K for 24 h. When cooled back to room temperature, the background pressure reached is typically at $3\times10^{-8}$ mbar.

Atmospheric compositions calculated for hot Jupiters using thermochemical equilibrium models constitutes a good starting point for the composition of the gas mixtures used in our experiments (Fleury et al. 2019). For this study, we used a gas mixture made of $D_2$ (Cambridge Isotope Laboratories, 99.99%), $D_2O$ (Alfa Aesar, 99.95%), and $^{13}CO$ (Cambridge Isotope Laboratories, 99.5%) with mixing ratio by volume of 99.26%, 0.48%, and 0.26%, corresponding to a mixture with a C/O ratio of 0.35. These are the three most abundant species predicted by thermochemical models for hot Jupiter atmospheres (excluding He, which is inert chemically) with a C/O ratio of 0.5 and T > 1000 K (Drummond et al. 2019; Moses et al. 2013; Venot et al. 2015). Carbon monoxide ($^{13}CO$), dihydrogen ($D_2$) and water ($D_2O$) were isotopically labelled to identify any contamination due to ambient atmospheric species. The gases were pre-mixed in a 2 L glass bulb. Then, the cell was filled with 15 mbar of the gas mixture at the room temperature and heated at 5 K min$^{-1}$ from room temperature (295 K) to various target temperatures: 1173 K, 1273 K, 1373 K, and 1473 K. The temperature of the tube was monitored at its center with three type-B thermocouples with equal spacing of 4.5 cm. After heating the cell to a desired temperature, the gas mixture was kept at that temperature for 21 h until the gas-phase reached a quasi-thermal-equilibrium composition. Evolution of the gas mixture composition was monitored during and after the heating using transmission Fourier Transform Infrared (FTIR) spectroscopy and mass spectrometry. Subsequently, the gaseous mixture was continuously irradiated for 24 h with vacuum ultraviolet (VUV, $\lambda < 200$ nm) photons generated by a microwave discharge lamp with a continuous 1.2 mbar flow of $H_2$ (Airgas, 99.9999%) powered with a microwave generator (OPTHOS) set to 70 W. It results in an intense emission at 121.6 nm (Ly$_\alpha$) and weaker emission in the 140-170 nm range (Ligterink et al. 2015). This reproduces the flux of UV photons received by hot Jupiters from their host stars with a predominance of Ly$_\alpha$ in the VUV (France et al. 2013; Miguel et al. 2015). Although only $H_2O$ can be directly photodissociated at these wavelengths, we demonstrated in our previous work that CO chemistry could be activated by photoexcitation (Fleury et al. 2019). The gases were kept at the desired temperature and irradiated for a similar amount of time to deconvolve the effects of thermochemistry from those of photochemistry. The evolution of the gas-phase composition was monitored using mass spectrometry during and after the irradiation. However, it was not possible to monitor the evolution of the gas-phase during the irradiation using IR spectroscopy because the CAAPSE setup has only two optical windows, which allow either VUV-irradiation or FTIR spectroscopy at any time, not simultaneously.

**Table 1: Summary of the pressure changes during the six heating experiments. Time listed is total time since filling the cell at room temperature with the $D_2$:$D_2O$:$^{13}CO$ gas mixture.**

|       | Initial Conditions (295 K) | | At the Thermal Equilibrium | | After UV Irradiation | |
|-------|----------|----------|----------|----------|----------|----------|
| T (K) | Time (h) | P (mbar) | Time (h) | P (mbar) | Time (h) | P (mbar) |
| 1173  | 0        | 15       | 24       | 28       | 48       | 28       |
| 1273  | 0        | 15       | 24       | 29       | 48       | -        |
| 1373  | 0        | 15       | 24       | 31       | 48       | 29       |
| 1473  | 0        | 15       | 24       | 32       | 48       | 28       |
| 1473  | 0        | 81       | **6**    | 170      | 207      | 94       |
| 1073  | 0        | 83       | **5**    | 140      | 206      | 140      |

The pressure in the cell was monitored at different steps of the experiments using a CDG-500 capacitance gauge (Agilent). The pressures measured for each studied temperature are summarized in Table 1. The heating of the gases involves an increase of the pressure, reflecting the thermal expansion of the gas. Evolution of the pressure in the cell during the irradiation is discussed further in Section 3.2.

## 2.2. Infrared Spectroscopy Analysis of the Gas Phase Composition

Evolution of the gas mixture composition was monitored with a Thermo Scientific Nicolet iG50 FTIR spectrometer. A collimated FTIR beam (a few mm in diameter) passed through our high temperature cell and was collected with a LN$_2$-cooled MCT-A detector. IR spectra were recorded in the 1500-6000 cm$^{-1}$ range with a resolution of 0.25 cm$^{-1}$ after a co-addition of 700 scans. The optical path length inside the cell is 48 ± 1 cm.

The concentration of the gaseous species detected with IR spectroscopy in this study are quantified using the Beer-Lambert law. The concentration of the absorbing molecules [C] (molecules cm$^{-3}$) is defined by the Equation (1):

$$[C] = \frac{\int_{\lambda_1}^{\lambda_2} A \, d\lambda}{l \times \int_{\lambda_1}^{\lambda_2} \sigma \, d\lambda} \quad (1)$$

where σ is the absorption cross-section (cm$^2$ molecule$^{-1}$) of the molecule at a given wavelength and the temperature T$_{max}$, $l$ is the path length of the beam through the cell, and $A$ is the absorbance at a given wavelength. $A$ and $\sigma$ are integrated over the absorption bands. As discussed in Fleury et al., (2019), for our calculation we assume that most of the gas is at the maximum temperature T$_{max}$. The cross-sections used were calculated for the temperature T$_{max}$ using the HITEMP and ExoMol databases (Rothman et al. 2010; Tennyson et al. 2016).

## 2.3. Mass Spectrometry Analysis of the Gas Phase Composition

The gas mixture composition was analyzed *in situ* with a Stanford Research System RGA200 quadrupole mass spectrometer (QMS) equipped with electron-multiplier, covering 1 to 200 *m/z* mass range with a resolution of 100 at *m/z* 100 (m/Δm). Gases were transferred to the QMS by opening a high-vacuum leak valve that separates the reaction gas-cell from the pumping system to which the QMS was attached. Pressure was kept to ~5×10$^{-7}$ mbar during all the measurements to enable comparison among the spectra. QMS ionization was achieved through electron impact at 70 eV.

## 2.4. Solid Phase Collection and Infrared Analysis of Thin Films

Similar to our previous studies (Fleury et al. 2019), the aerosol samples were produced during two longer experiments, at a higher pressure of 81 mbar and at 1073 K and 1473 K respectively. After heating the cell to a desired temperature, the gas mixture was kept at that temperature for 2 h and subsequently irradiated for 201 h. The samples were collected on two sapphire substrates (25 mm diameter and 1 mm thick) placed inside CAAPSE. After the irradiation, the temperature was ramped down to room temperature and the volatiles were pumped off. Subsequently, the cell was opened to ambient air and the samples were transferred to the FTIR for analysis. Transmission infrared spectra of the samples were measured with a Thermo Scientific Nicolet 6700 FTIR Spectrometer. The infrared signal was collected by a Deuterium TriGlycine Sulfate (DTGS) detector in the 1600 cm$^{-1}$ (sapphire window absorption limit) to 4000 cm$^{-1}$ range with a resolution of 1 cm$^{-1}$ after a co-addition of 300 scans.

## 3. Results

### 3.1. Thermochemistry in $D_2$/$D_2O$/$^{13}CO$ Gas Mixtures

As a control, we investigated the effect of the heating of the cell without UV irradiation on the gas mixture composition. Figure 1 presents the IR spectra of the gas mixture at ambient temperature (~295 K) and after 21 h of heating once the different set oven temperatures (1173 K, 1273 K, 1373 K, and 1473 K) were reached.

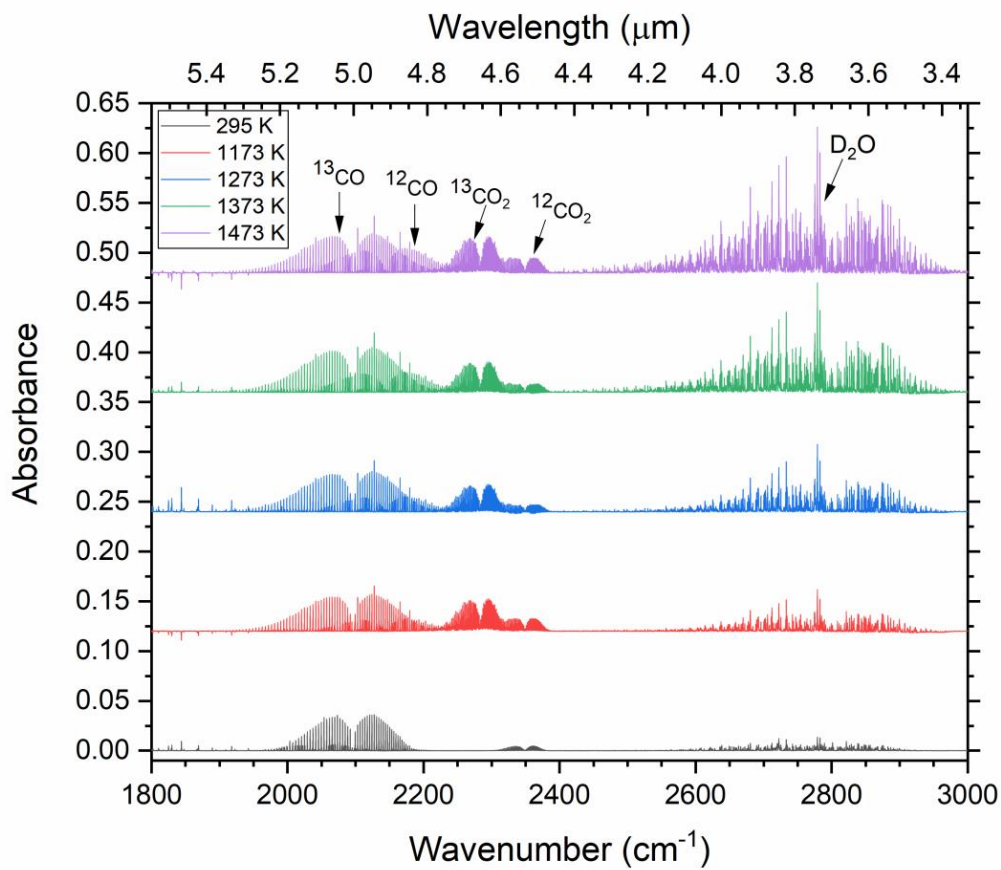

**Figure 1: IR spectra of the initial gas mixture of $D_2$:$D_2O$:$^{13}CO$ (99.26:0.48:0.26) at 295 K and after 21 h of heating at different temperatures: 1173 K, 1273 K, 1373 K, and 1473 K.**

The absorption bands of $^{13}$CO at 2095 cm$^{-1}$, D$_2$O centered at 2780 cm$^{-1}$, and $^{12}$CO$_2$ at 2349 cm$^{-1}$ are visible on the spectrum recorded at room temperature (lowest trace of Figure 1). The absorption bands of $^{12}$CO$_2$ can be attributed to the signature of the residual air present on the optical pathway outside of the cell. For all studied temperatures, after 21 h of heating we observed a new absorption band at 2282 cm$^{-1}$ attributed to the formation of $^{13}$CO$_2$ as well as an increase of the D$_2$O absorbance correlated with the increase of the temperature, highlighting an efficient formation of water in these conditions. Although thermochemistry in our experiments is not yet very well understood, thermochemical reactions are likely responsible for the production of these two species ($^{13}$CO$_2$ and D$_2$O). In addition, a small increase of $^{12}$CO$_2$ absorbance as well as the detection of the new bands around 2146 cm$^{-1}$ attributed to $^{12}$CO, have been observed. The origin of these contaminations is not yet fully understood but several scenarios (such as outgassing of the alumina tube with $^{12}$C-containing organic residue) were discussed in our previous work. In a control experiment, we have observed that gas released from the tube was dominated by H$_2$O, $^{12}$CO, and $^{12}$CO$_2$ (Fleury et al. 2019). It is therefore likely that natural isotope molecules observed in our experiment such as $^{12}$CO is due to outgassing from the tube during the experiments. Those species were found to have a negligible impact on the simulated chemistry because they were the same species as the ones used in the initial gas mixtures but with different isotopes (Fleury et al. 2019), and a similar analysis can be used for the present work.

Additionally, we analyzed the gas phase composition using mass spectrometry. Figure 2 (top) presents the mass spectra of the gas phase for *m/z* 1 to 50 at 295 K and at the different studied temperatures. Spectra are displayed only up to *m/z* 50 as no other peaks were observed at higher masses. To facilitate the visualization of the smallest mass peaks, Figure 2 (bottom) presents the same mass spectra from *m/z* 15 to 50. Spectra are normalized to the deuterium peak (*m/z* 4) to allow quantitative comparison of the different spectra. We have chosen to use the deuterium peak because its concentration did not change significantly during the heating of the gases. A list of the mass peaks observed in the spectra are shown in Figure 2 and their assignments are given in Table 2.

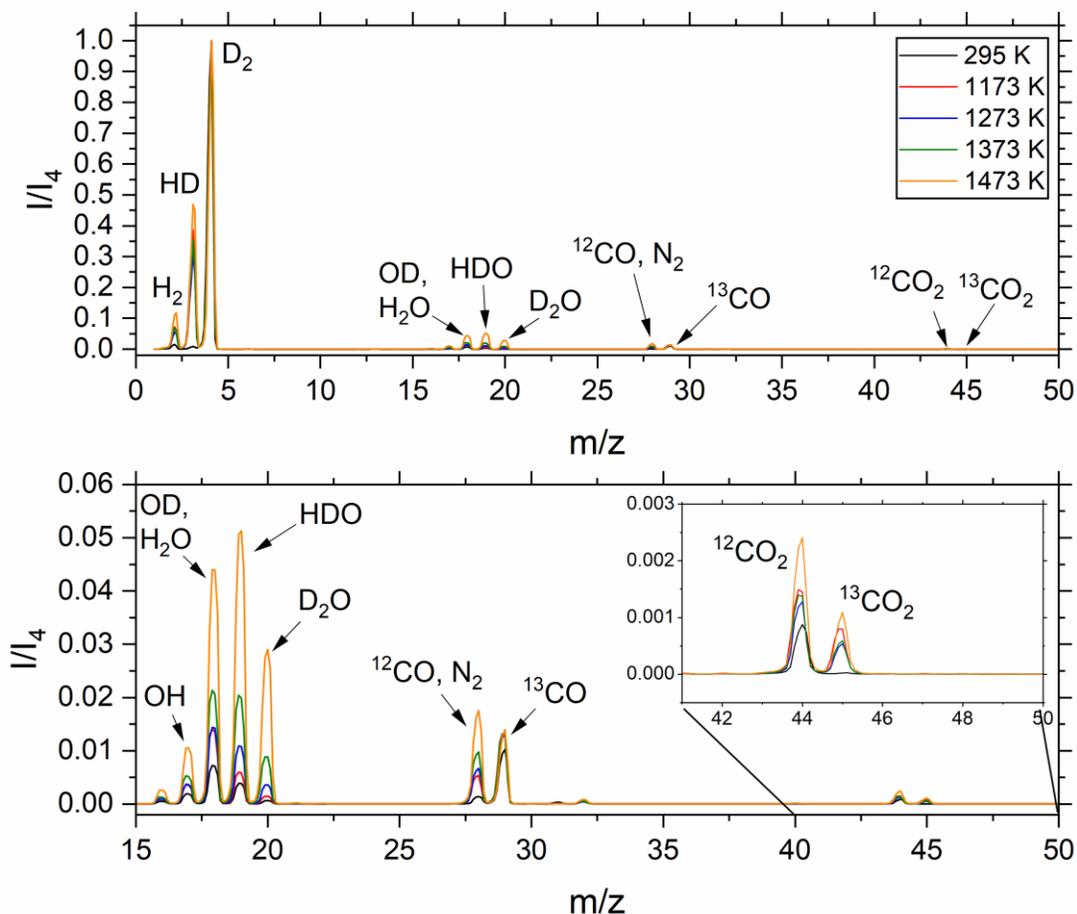

**Figure 2: Mass spectra from *m/z* 1 to 50 of the gas mixture at ambient temperature (295 K) and after 21 h of heating at different set oven temperatures: 1173 K, 1273 K, 1373 K, and 1473 K (top). Same mass spectra of the gas from *m/z* 15 to 50 (bottom). Mass spectra are normalized to the D$_2$ peak intensity at *m/z* 4.**

In general, the results obtained with mass spectrometry are in good agreement with the ones obtained with IR spectroscopy. We observed an increase of the intensity at *m/z* 45 attributed to $^{13}CO_2^+$. Moreover, we observed an increase of the intensity at *m/z* 20, 19 and 18 attributed to $D_2O^+$, $HDO^+$, and $OD^+$ (fragment of $D_2O$) respectively, although a contribution of $H_2O^+$ at *m/z* 18 cannot be totally ruled out with this dataset. In addition, we observed an increase of the peaks at *m/z* 2 and 3, which can be attributed to $H_2^+$ and $HD^+$. The increase of $H_2$ can be explained by a release of adsorbed $H_2$ molecules during the heating of the gases while the HD can be explained by isotope exchanges between the released or adsorbed $H_2$ and $D_2$.

**Table 2: Assignments of mass peaks observed in the mass spectra show in Figure 2, 4, and 5.**

| Peak (*m/z*) | Species | Peak (*m/z*) | Species | Peak (*m/z*) | Species |
|---|---|---|---|---|---|

| 2  | $H_2^+$    | 16 | $O^+$           | 29 | $^{13}CO^+$   |
|----|------------|----|-----------------|----|---------------|
| 3  | $HD^+$     | 16 | $O^+$           | 32 | $O_2^+$       |
| 4  | $D_2^+$    | 17 | $OH^+$          | 40 | $Ar^+$        |
| 12 | $^{12}C^+$ | 18 | $OD^+, H_2O^+$  | 44 | $^{12}CO_2^+$ |
| 13 | $^{13}C^+$ | 20 | $D_2O^+$        | 45 | $^{13}CO_2^+$ |
| 14 | $N^+$      | 28 | $^{12}CO^+, N_2^+$ |  |            |

Finally, we quantified the changes in mixing ratios of $^{13}CO$, $^{13}CO_2$, and $D_2O$ at ambient temperature at the beginning of the experiments vs. after 21 h of heating at the different studied temperatures using the method described in Section 2.2. These ratios are summarized in Table 3. For the all the studied temperatures, the $^{13}CO_2$ mixing ratio was between 240 and 360 ppm$_v$. Despite the significant formation of $CO_2$, mass spectrometry results show that CO remained the major carbonaceous species for all the experiments. However, the relative abundance of CO could not be quantified at high temperature from the IR spectra because the $^{13}CO$ absorption band overlapped with the one attributed to $^{12}CO$.

**Table 3: Mixing ratio of $^{13}CO$, $^{13}CO_2$, and $D_2O$ calculated from IR spectra at 295 K, after 21 h of heating for different set oven temperatures: 1173 K, 1273 K, 1373 K, and 1473 K, and after subsequent 24 h of UV irradiation. The uncertainties are given at 2 standard deviations and were calculated from the standard fluctuations of the infrared spectroscopy measurements.[a] Fleury et al., (2019).**

|  |  | C/O = 0.35 | | | C/O = 1 |
|---|---|---|---|---|---|
|  | T (K) | $^{13}CO$ | $^{13}CO_2$ | $D_2O$ | $^{13}CO_2$[a] |
| Initial Conditions | 295 | $(2.6 \pm 0.2) \times 10^{-3}$ | - | $(4.8 \pm 0.3) \times 10^{-3}$ | - |
| After Heating | 1173 | - | $(3.5 \pm 0.2) \times 10^{-4}$ | $(7.3 \pm 0.4) \times 10^{-3}$ | $(1.4 \pm 0.1) \times 10^{-4}$ |
|  | 1273 | - | $(2.4 \pm 0.2) \times 10^{-4}$ | $(1.3 \pm 0.7) \times 10^{-2}$ | $(1.8 \pm 0.2) \times 10^{-4}$ |
|  | 1373 | - | $(2.6 \pm 0.2) \times 10^{-4}$ | $(2.0 \pm 0.1) \times 10^{-2}$ | $(8.0 \pm 0.4) \times 10^{-5}$ |
|  | 1473 | - | $(3.6 \pm 0.2) \times 10^{-4}$ | $(3.3 \pm 0.2) \times 10^{-2}$ | $(3.4 \pm 0.2) \times 10^{-5}$ |
| After UV Irradiation | 1173 | - | $(8.8 \pm 0.5) \times 10^{-4}$ | $(7.3 \pm 0.4) \times 10^{-3}$ | $(1.2 \pm 0.1) \times 10^{-4}$ |
|  | 1273 | - | $(8.9 \pm 0.5) \times 10^{-4}$ | $(1.3 \pm 0.7) \times 10^{-2}$ | $(1.0 \pm 0.1) \times 10^{-4}$ |
|  | 1373 | - | $(6.0 \pm 0.3) \times 10^{-4}$ | $(2.3 \pm 0.1) \times 10^{-2}$ | $(6.4 \pm 0.1) \times 10^{-4}$ |
|  | 1473 | - | $(5.6 \pm 0.3) \times 10^{-4}$ | $(3.9 \pm 0.2) \times 10^{-2}$ | $(4.3 \pm 0.1) \times 10^{-4}$ |

Despite the similarities in $^{13}CO_2$ mixing ratios, the concentration of $D_2O$ was shown to increase regularly with temperature from the initial 0.48% at 295 K to 3% at 1473 K. These results differ quantitatively from the ones obtained in our previous study with a gas mixture made of $H_2$ and $^{13}CO$ only (Fleury et al. 2019). In the previous study, in the absence of water, the amount of carbon dioxide produced by thermochemistry was 2 to 10 times lower for all studied temperatures, and increased to a maximum at 1273 K before decreasing at higher temperatures. In our current study, the amount of $^{13}CO_2$ produced is similar for all studied temperatures. More interestingly, no production of water was observed in our previous study (or lower than the natural variation of the water content on the optical pathways outside of the cell) under thermal equilibrium only, while the amount of water produced in this study varies from 0.7% at 1173 K up to 3.3% at 1473 K. In summary, more $^{13}CO$ is thermochemically converted to $^{13}CO_2$ and $D_2O$ when water is initially added to the gas mixture to reduce the C/O ratio to 0.35. Finally, we do not observe the formation of methane in any of these experiments, while we have observed the formation of ~40 to 80 ppm$_v$ of $CH_4$ in our previous experiments with a gas mixture with a C/O ratio of 1 (Fleury et al. 2019). Although thermochemistry in our experiment is not fully understood, the inhibition of $CH_4$ production appears to be correlated to the increase of $CO_2$ production. Indeed, we have observed in this study an increase of the $CO_2$ production (Table 3) for all temperatures compared to Fleury et al., (2019). It is likely that the formation of $CO_2$ and $CH_4$ from the initial CO in our experiments result from two competitive chemical pathways. $CO_2$ is the main product for both C/O ratios studied (i.e. 0.35 and 1), but our results show that in gas mixtures with lower C/O ratio, the formation of $CO_2$ is enhanced while the formation of $CH_4$ is inhibited.

### 3.2. UV Irradiation of Gas Mixtures at Thermal Equilibrium

After 21 h of heating, the same gas mixtures were irradiated with UV photons (Ly$_\alpha$) to simulate photochemistry in hot Jupiter atmospheres. Figure 3 presents the IR spectra of the gas mixtures after 24 h of irradiation for the different set oven temperatures: 1173 K, 1273 K, 1373 K, and 1473 K. The spectra show the absorption bands of carbon monoxide ($^{12}CO$, and $^{13}CO$), carbon dioxide ($^{12}CO_2$, and $^{13}CO_2$) and $D_2O$. Mixing ratios of $^{13}CO_2$ and $D_2O$ were calculated for the four temperatures and these values are summarized in Table 3. Since $^{12}CO$ and $^{13}CO$ absorption bands overlapped, we could not perform a quantification of $^{13}CO$.

For all studied temperatures, the mixing ratio of $^{13}CO_2$ increased after irradiation compared to heating only: by a factor of 3 at 1173 K and 1273 K, and by a factor of 2 at 1373 K and 1473 K. In Fleury et al., (2019), we explained the enhancement of $CO_2$ after irradiation by the photoexcitation of CO by Ly$_\alpha$ photons followed by the reactions of the excited molecules with ground-state CO. Similar reactions could explain the increase of the $CO_2$ concentration. On the contrary, the mixing ratio of $D_2O$ did not vary significantly after irradiation, although water should be significantly dissociated by UV photons emitted by the hydrogen lamp. It implies that a competitive mechanism is responsible for efficiently recycling water, leading to a good stability of water vapor in these conditions. In previous works, we have proposed that water could be efficiently formed photochemically through the reaction of photochemically produced O ($^1$D) radicals with $D_2$ to form OD and finally $D_2O$ (Fleury et al. 2015, 2019). Here, water photodissociation by $Ly_\alpha$ photons would result in the formation of OD and D radicals. OD can react directly with excess of dihydrogen in the gas phase leading to the reformation of a water molecule. These results are in good agreement with different chemical models which have also predicted an efficient recycling of water in hot Jupiter atmospheres despite an efficient destruction by UV photons through the same mechanisms (Line et al. 2011, 2010; Moses et al. 2011).

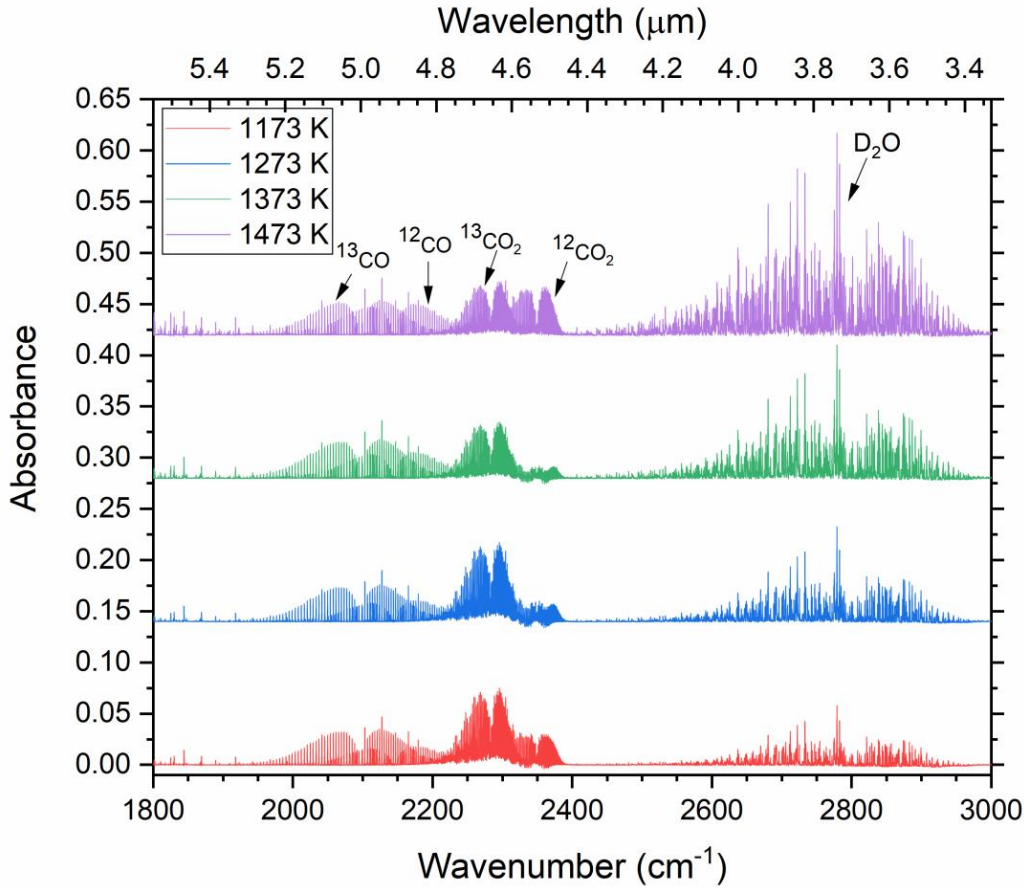

**Figure 3: IR spectra of the gas mixture after 24 h of irradiation with UV photons ($Ly_\alpha$) for the different studied temperatures: 1173 K, 1273 K, 1373 K, and 1473 K.**

To complete the analysis of the gas phase composition, we monitored its evolution upon UV irradiation using mass spectrometry. Figure 4 presents the mass spectra of the gas phase for *m/z* 1 to 50 at the different studied temperatures after 24 h of irradiation.

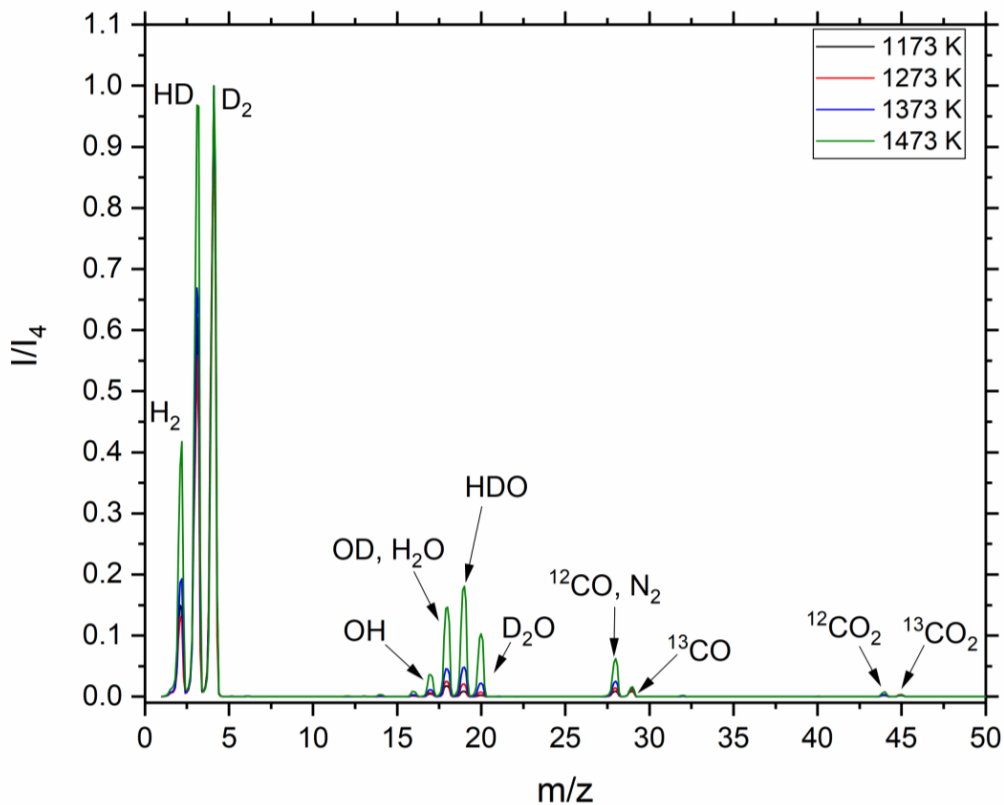

**Figure 4: Mass spectra from *m/z* 1 to 50 of the gas mixture after 24 h of UV irradiation at different set oven temperatures: 1173 K, 1273 K, 1373 K, and 1473 K. Mass spectra are normalized to the D$_2$ peak intensity at *m/z* 4.**

In addition, Figure 5 presents the evolution of the mass peak intensities after 24 h of irradiation compared to the mass peak intensities after 21 h of heating ($I_{irradiation} - I_{heating}$) for the different set oven temperatures. Figure 5 presents mass spectra only from *m/z* 10 to 50 to facilitate the visualization of the smallest mass peaks.

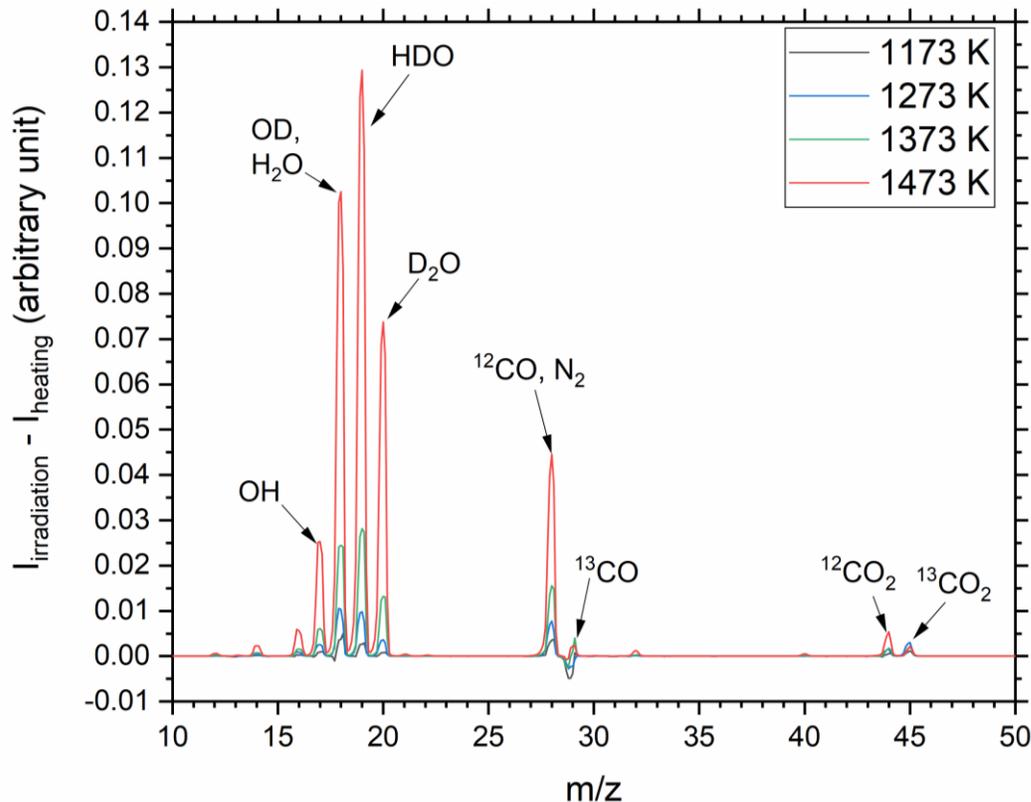

**Figure 5: Evolution of the mass peak intensities after 24 h of irradiation compared to the mass peak intensities after 21 h of heating ($I_{irradiation} - I_{heating}$) for the different set oven temperatures: 1173 K, 1273 K, 1373 K, and 1473 K, from *m/z* 10 to 50.**

The intensity of most mass peaks including $H_2$, HD, $^{13}CO_2$ and $D_2O$ increase as a function of the temperature relative to $D_2$ in contradiction with quantifications made from IR spectroscopy data. Because these spectra are normalized to *m/z* 4 ($D_2$), it can be explained by a significant decrease of the $D_2$ amount in the gas mixtures during the irradiation leading to the increase of the relative intensity for all the other mass peaks. This is presumably confirmed by the decrease of the intensity of the peak at *m/z* 4 after irradiation in the non-normalized spectra although an absolute quantification is not possible. This hypothesis is also supported by the measures of pressure made before and after irradiation at the studied temperatures. These values are summarized in Table 1. While at 1173 K, the pressure did not change during the irradiation; at high temperature, we observed a decrease of the pressure of 2 mbar at 1373 K and 3 mbar at 1473 K. This decrease of the pressure associated with the decrease of the $D_2$ mixing ratio points towards the formation of more complex molecules resulting in the decrease of the total number of molecules in the gas phase and of the pressure. The only deuterated molecule observed is $D_2O$. However, its mixing ratio was constant during the irradiation and cannot be the product of the photochemistry. If another gaseous product was produced during the irradiation, it was not detected using IR spectroscopy or mass spectrometry. Another explanation could be the formation of a solid product. In our previous study with a gas mixture made of $H_2$ and CO, we identified the photochemical formation of an organic solid polymer made of C, O, and H, at 1473 K (Fleury et al. 2019).

### 3.3. Photochemical Formation of Refractory Solid Organics

For the photochemical experiments made at 1373 K and 1473 K presented above, we observed a decrease of the pressure at the end of the irradiation, likely due to the formation of more complex molecules. Although a detectable amount of solid-phase material was not observed in these results, a similar trend in the pressure observed in our previous work (Fleury et al. 2019) was associated with the formation of solid organic aerosols. To investigate the possibility of organic aerosol formation, we repeated the 1473 K experiment at a higher pressure of 81 mbar (measured at 295 K) and for a longer irradiation of 207 h to have a larger number of molecules to convert into a solid phase.

After the heating of the gas phase to 1473 K, the pressure in the cell was 170 mbar (Table 1) and we observed a decrease of the pressure after irradiation to 94 mbar, highlighting the conversion of a part of the initial gas mixture into more complex molecules. However, no solid deposits were visible on the sapphire substrates after the experiments and the analysis of the substrates with transmission infrared spectroscopy did not reveal the presence of any absorption bands attributable to a solid-phase. In our previous study, solid organic products were observed as thin films deposited on the substrates while here no products were observed. The formation mechanism of these thin films in not yet fully known but they presumably grow from reactive gas species present in the cell via the formation of nanometer-size monomers, which finally aggregate. Then the particles grow by deposition of species present in the gaseous phase and deposit as thin films. Our results suggest that in low C/O gas mixtures, the critical number density of monomers is not reached to start the growth of larger particles because of the low reactivity of CO and $CO_2$ or that the amount of aerosol produced is significantly decreased, below the limit of detection of our infrared spectrometer.

However, during the cooling of the gas mixtures after the experiments, we have observed the deposition of a residue on the flanges and the $MgF_2$ windows. The formation of this solid residue is observed only on the coolest part of the cell during the cooling of the gases and not on the sapphire windows disposed in the center of the cell where the temperature is higher. Figure 6 presents the IR absorption spectrum of the gas mixture after 201 h of irradiation at 1473 K and the subsequent cool down to room temperature. In addition of the absorption bands of $^{12}CO$, $^{13}CO$, $^{12}CO_2$, $^{13}CO_2$, and $D_2O$, in the gas-phase, we observed two broad absorption bands centered at ~2400 cm$^{-1}$ and 3300 cm$^{-1}$. These two bands can be attributed to the solid residue deposited on the MgF$_2$ windows during the cooling of the gases. We assign these bands to hydroxyl -OD and -OH groups of organic molecules, indicating that this solid residue should be low-molecular weight oxidized organics that are solids at room temperature but volatile at the higher temperature (1473 K) used in our experiments to simulate hot Jupiter atmospheres. The position and shapes of these bands differ from the ones observed in our previous experiments (Fleury et al. 2019), indicating different formation processes. It is likely that the formation of these volatile organic molecules, more complex than the initial $D_2$, $^{13}CO$, and $D_2O$ molecules, is responsible for the decrease of the pressure in the chamber observed after the irradiation of the gas mixture. Although this detection is important for the understanding of the chemical processes in our experiments, this solid formation cannot be directly connected to any process occurring in hot Jupiter atmosphere, because of the lower temperature at which the solid formation has been observed.

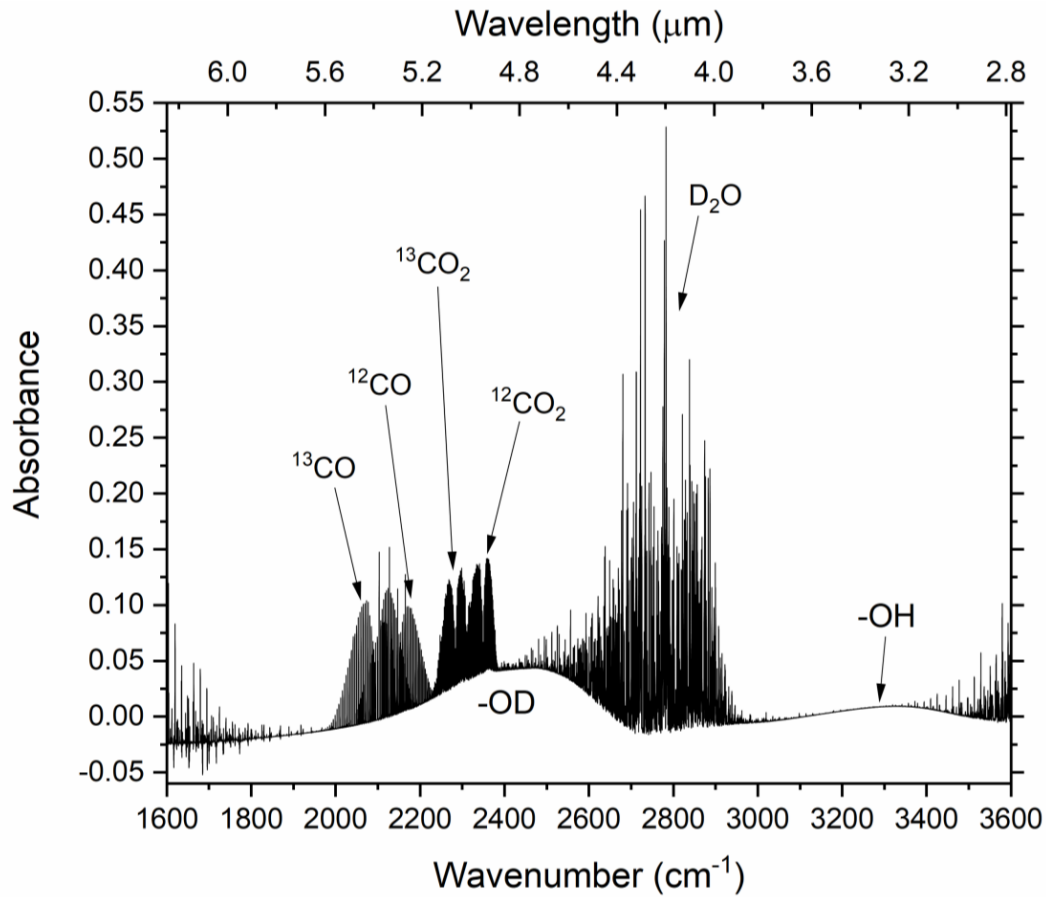

**Figure 6: IR spectrum of the gas mixture after 201 h of irradiation with UV photons (Ly$_\alpha$) at 1473 K and subsequent cool down to room temperature.**

To explore further the question of the photochemical formation of aerosols in low C/O ratio atmospheres, we repeated this experiment at a lower temperature of 1073 K. At this temperature we did not observe any decrease of the pressure after irradiation and we did not observe the formation of any solid deposit in the experiments. These results show that at lower temperature, the conversion of simple gas molecules to more complex species is drastically less efficient than at higher temperature. At higher temperature, reactivity may be favored by faster kinetics or an increase of the photolysis rates due to the increase of the absorption cross-section of molecules with the temperature (Venot et al. 2018).

## 4. Implications for Hot Jupiter Atmospheres

Our experimental results demonstrate that the C/O ratio and the molecular compositions of hot Jupiter atmospheres strongly affect their thermochemistry and photochemistry. On the one hand, our previous experiment with a $H_2$:CO atmosphere (C/O = 1) thermochemistry led to the formation of 16 to 180 ppm$_v$ of carbon dioxide as a function of the temperature, while in the present study with a H$_2$:CO:H$_2$O atmosphere (C/O = 0.35) we observed the production of 240 to 360 ppm$_v$, highlighting that the thermochemical production of CO$_2$ is less efficient and more temperature dependent in high C/O ratio hot atmospheres. On the other hand, our studies also show that photochemistry can enhance the CO$_2$ production by a factor of 10 in high C/O atmospheres while in low C/O atmospheres, we observe only an enhancement by a factor of 2 to 3. In addition, despite a shorter time of irradiation (18 h vs 24 h in the present work) the mixing ratios of CO$_2$ were similar after irradiation for both studies for all studied temperatures. These results indicate that the C/O ratio and the gas mixture composition strongly influence the pathways responsible for the formation of CO$_2$ in hot Jupiter atmosphere. Thermochemical reactions are primarily responsible for the formation of CO$_2$ in low C/O ratio atmospheres, while photochemistry is the major process responsible for the formation of CO$_2$ in high C/O ratio atmospheres. However, in both cases, the relative amount of CO$_2$ produced at high temperature and under UV irradiation is similar at a given temperature highlighting that the carbon dioxide concentration may not differ significantly between hot Jupiter atmospheres with a low or a high C/O ratio, although we will need to explore a wider range of compositions to firmly conclude on the impact of the C/O ratio on the CO$_2$ mixing ratio.

Second, our studies point out that water chemistry can significantly differ as a function of the C/O ratio. In the H$_2$:CO gas mixture, thermochemical production of water was found to be inefficient while photochemistry was shown to efficiently promote the formation of water with a production of water increasing with the temperature. On the contrary, the D$_2$:D$_2$O:$^{13}$CO gas mixture thermochemistry led to efficient water production with mixing ratios ranging from 0.7% at 1173 K to 3.3% at 1473 K. Photochemistry driven by the VUV radiation does not modify further the water mixing ratio for the low C/O (present study), while in H$_2$:$^{13}$CO (C/O = 1) experiments photochemical production of water increases with the temperature. Chemical models predict that water abundances vary with the C/O ratio: atmospheres with a low C/O ratio being water-rich and atmospheres with a high C/O ratio being water-poor (Drummond et al. 2019; Goyal et al. 2018; Heng & Lyons 2016; Moses et al. 2013; Tsai et al. 2017; Venot et al. 2015). Though our results are in good agreement with these theoretical studies when considering hot Jupiter atmospheres at the thermal equilibrium, our results suggest that disequilibrium chemistry such as the VUV photochemistry could drastically affect the water mixing ratio in atmospheres with a high C/O ratio and at high temperatures. In these cases, the use of the H$_2$O mixing ratio to estimate the planetary C/O ratio could lead to biased results. In addition, our results demonstrate that UV photolysis does not affect the abundance of water in low C/O ratio atmospheres because H$_2$O is efficiently recycled in agreement with model calculations (Line et al. 2011, 2010; Moses et al. 2011). Then, a low water abundance derived may not be the result of disequilibrium chemistry, but would rather reflect a high C/O ratio in hot Jupiters with lower temperatures or would reflect the presence of additional opacities such as those from aerosols.

Finally, our experimental results show that the C/O ratio and the molecular composition of hot Jupiter atmospheres can drastically affect the formation of photochemical organic aerosol. In H$_2$:CO gas mixture, we have observed the formation of a solid organic product after irradiation at 1473 K (Fleury et al. 2019) while no solid-phase production is observed at 1473 K after the irradiation of D$_2$:D$_2$O:$^{13}$CO, demonstrating that the organic growth and the aerosol production is inhibited in low C/O ratio atmospheres with a high water mixing ratio. Our experimental results suggest that the presence of photochemical organic aerosols may be possible only in hot Jupiter atmospheres with a high C/O ratio, although other aerosols **may** be present in atmospheres with a low C/O ratio such as clouds or sulfur aerosols. Although the process of formation of aerosols in hot Jupiter atmospheres is poorly understood, the limitation of the organic growth in low C/O atmospheres could be explained by the formation of shorter hydrocarbon chains, because of the reactions of hydrocarbons with H$_2$O or radicals such as OH and O reduce the probability of high-molecular-weight hydrocarbons. Therefore, if transit spectra of hot Jupiter atmospheres present the spectral signature of photochemical organic aerosols, it could be interpreted as an indicator that these atmospheres have a high C/O ratio. If we combine these experimental findings with the observational determination that most host stars have lower C/O ratios than the Sun (i.e. C/O ratio < 0.54) (Brewer et al. 2017), then it implies that we should find photochemical organic aerosols only in planets presenting a carbon enrichment compared to their host stars (with high C/O ratios).

## Conclusion

We have conducted a new experiment to study the chemistry and formation of aerosols in hot Jupiters with a C/O ratio < 1. We have irradiated D$_2$:$^{13}$CO:D$_2$O gas mixtures at various temperatures from 1200 K to 1500 K with Ly-α (121.6 nm) photons to reproduce photochemistry in these hot atmospheres and we have monitored the evolution of the gas-phase composition using infrared spectroscopy and mass spectrometry. Finally, we have compared these results with the ones obtained previously experimentally in Fleury et al., (2019) for hot Jupiter atmospheres with a higher C/O ratio of 1, to assess the role of the C/O ratio on hot Jupiter atmospheric compositions.

First, we have demonstrated that thermochemistry led to a significant formation of carbon dioxide and water at each studied temperature. After 21 h of heating, similar mixing ratios of a few hundred $ppm_v$ were quantified for $CO_2$ for all studied temperatures. In contrary, the water mixing ratio was found to increase with the temperature up to 3.3% at 1473 K.

Second, we subsequently irradiated these equilibrium gas mixtures with UV photons. Photochemistry was found to promote a limited increase of the $CO_2$ mixing ratio by a factor of 2-3. Comparison with our previous study demonstrated that the C/O ratio and the gas mixture composition strongly influence the pathways responsible for the formation of $CO_2$ in hot Jupiter atmosphere. Our study shows that thermochemical reactions are primarily responsible for the formation of $CO_2$ in low C/O ratio atmospheres, while photochemistry is the major process responsible for the formation of $CO_2$ in high C/O ratio atmospheres. However, in both cases, the relative amount of $CO_2$ produced at high temperatures and under UV irradiation is similar at a given temperature.

In addition, the $H_2O$ mixing ratio was found to not change significantly under UV irradiation. This result highlights that water is very stable in hot Jupiter atmospheres despite its efficient dissociation by UV photons. Indeed, water recycling is very efficient in dihydrogen-dominated atmospheres because OH radicals produced by $H_2O$ photodissociation can react with $H_2$ to reform $H_2O$.

Finally, the formation of a detectable amount of non-volatile solid organic thin films was not observed after irradiation of the gas mixtures at 1473 K and 1073 K with $Ly_\alpha$ photons. This result demonstrates that the C/O ratio and the initial gas composition significantly affects the efficiency of the aerosol formation. For atmospheres with a C/O ratio < 1, water and its dissociation products (OH and O) likely inhibit the growth of organic molecules and the formation of aerosols, suggesting that photochemical organic aerosols are likely to be observed in planets presenting a carbon enrichment compared to their host stars.

# Acknowledgments

This research has been carried out at the Jet Propulsion Laboratory, California Institute of Technology, under a contract with the National Aeronautics and Space Administration. This work was supported by the JPL Strategic R&TD funding under the "Exoplanet Science Initiative, ESI" and by the NASA Exoplanet Research Program. © 2020 California Institute of Technology. All right reserved